\documentclass[12pt]{article}

\usepackage{scicite}

\usepackage{amssymb}
\usepackage{times}
\usepackage{amsmath}
\usepackage[T1]{fontenc} 
\usepackage[utf8]{inputenc} 
\usepackage{float} 
\usepackage{graphicx} 
\usepackage{url}

\newenvironment{sciabstract}{%
\begin{quote} \bf}
{\end{quote}}

\newcounter{lastnote}

\title{Photochemical CS$_2$ Gas Detected on a 20-Myr-old Exoplanet}

\author
{Fei Dai,$^{1\ast}$ Erik Petigura,$^{2}$ John Livingston,$^{3,4,5}$ Nicholas Wogan, $^{6}$\\
Sagnick Mukherjee, $^{7}$ Zhecheng Hu,$^{8}$ Ian J.\ M.\ Crossfield, $^{9}$ \\
James Owen, $^{10}$ Kento Masuda $^{11}$\\
\normalsize{$^{1}$ Institute for Astronomy, University of Hawai`i at Mānoa, 2680 Woodlawn Dr, Honolulu, HI, 96822, USA}\\
\normalsize{$^{2}$Department of Physics \& Astronomy, University of California Los Angeles, Los Angeles, CA, USA}\\
\normalsize{$^{3}$Astrobiology Center, 2-21-1 Osawa, Mitaka, Tokyo, 181-8588, Japan}\\
\normalsize{$^{4}$National Astronomical Observatory of Japan, 2-21-1 Osawa, Mitaka, Tokyo, 181-8588, Japan}\\
\normalsize{$^{5}$Department of Astronomical Science, The Graduate University for Advanced Studies, Japan}\\
\normalsize{$^{6}$NASA Ames Research Center, Moffett Field, CA 94035}\\
\normalsize{$^{7}$School of Earth and Space Exploration, Arizona State University, Tempe, AZ, USA}\\
\normalsize{$^{8}$Department of Astronomy, Tsinghua University, Beijing 10084, People’s Republic of China}\\
\normalsize{$^{9}$Department of Physics and Astronomy, University of Kansas, Lawrence, KS, USA}\\
\normalsize{$^{10}$Astrophysics Group, Department of Physics, Imperial College London, London, UK}\\
\normalsize{$^{11}$Department of Earth and Space Science, Osaka University, Osaka, Japan}\\
\normalsize{$^\ast$To whom correspondence should be addressed; E-mail:  fdai@hawaii.edu.}
}

\date{}

\begin{document}

\baselineskip24pt

\maketitle

\begin{sciabstract}
Probing the atmospheres of young exoplanets offers a powerful window into how planetary systems evolve and the physical and chemical processes that drive those early evolutions. We present JWST/NIRSpec transmission spectroscopy of V1298 Tau e, a $\sim$20-Myr-old, $\sim$15-$M_\oplus$ planet with a Jupiter-like radius orbiting a young Sun-like star. We identified carbon disulfide (CS$_2$) in its atmosphere at $>$8$\sigma$ significance based on spectral features between  4.3 and 4.7~$\mu$m. Photochemical forward models show that the inferred CS$_2$ abundance is physically plausible in an H/He-dominated atmosphere exposed to intense ultraviolet irradiation. The atmosphere of V1298 Tau e is strikingly different from its nearest neighboring planet b, whose atmosphere shows SO$_2$ rather than CS$_2$. These observations demonstrate that even planets within the same system can occupy distinct photochemical regimes. Our results further provide empirical evidence for complex sulfur photochemistry in exoplanet atmospheres in general and may also point to divergent formation or evolutionary pathways within the same planetary system.
\end{sciabstract}

\bigskip

\noindent {\bf Main Text:} 

V1298 Tau is a young ($23\pm4$ Myr-old) T-Tauri star in the Taurus--Auriga association that is slightly more massive than the Sun (1.1~$M_{\odot}$ \cite{David2019}). It hosts four transiting planets with periods of 8.2, 12.4, 24.1, and 48.7 days and sizes ranging from 5 to 10 $R_\oplus$. The planets are near mean-motion resonances (with period ratios close to 3:2--2:1--2:1) and exhibit large transit-timing variations \cite{Agol} that enabled dynamical mass measurements of $\sim$5--15 $M_\oplus$ \cite{David_4p,Livingston,Hu2025}. The outermost planet, V1298 Tau e, has a Jupiter-like radius of $\sim$10 $R_\oplus$ but a mass of only $\sim$15 $M_\oplus$, making it an inflated progenitor of a mature sub-Neptune, the most common type of exoplanets discovered in our Galaxy to date\cite{Fressin,Petigura2013,Zhu}. V1298 Tau e provides a rare view of a young proto-sub-Neptune, when its extended envelope has not yet fully contracted \cite{Fortney2007} or been eroded \cite{Owen2013,Lopez2014}, and when intense high-energy irradiation from the host star drives vigorous photochemistry \cite{Zahnle2009,Moses,Venot,Hu2012,Tsai2023}.

\begin{figure}[t!]
\begin{center}
\includegraphics[width=\textwidth]{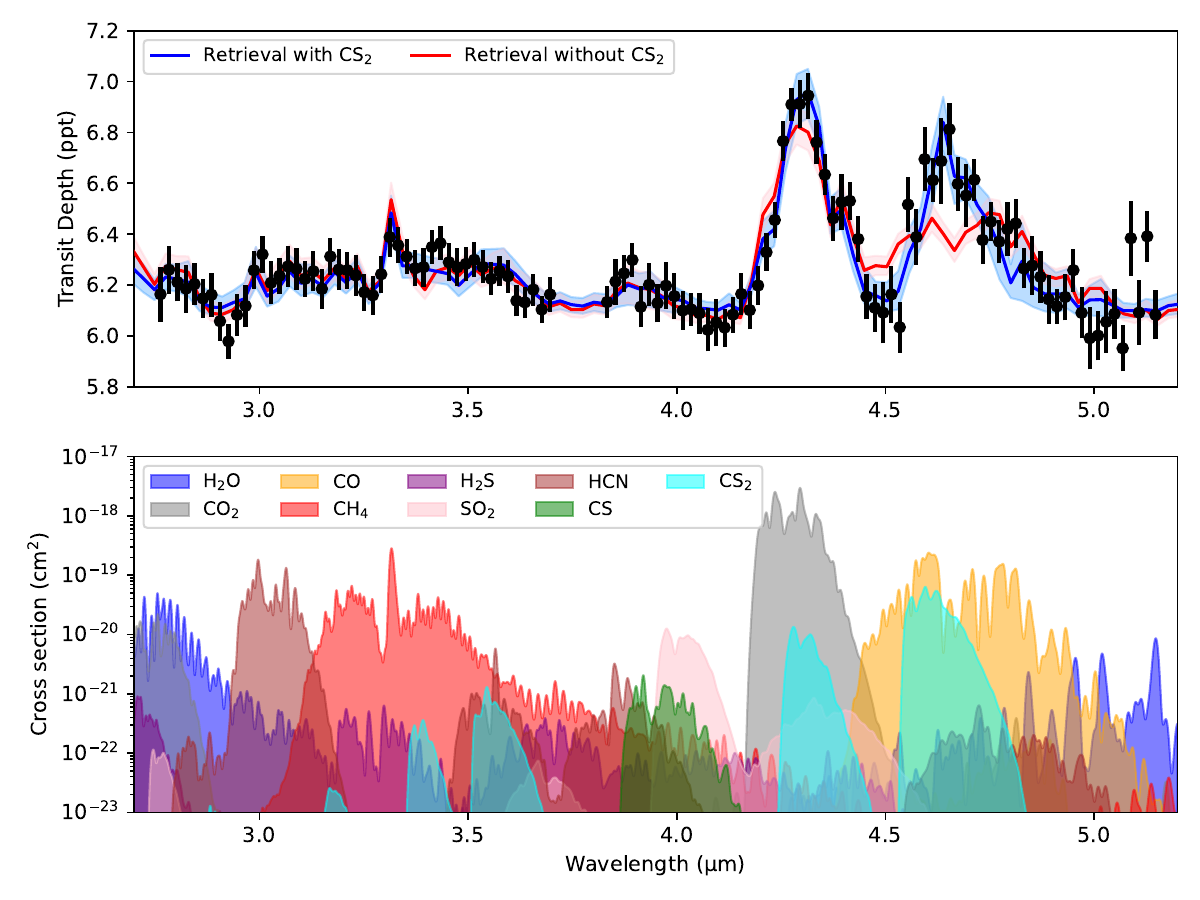}
\end{center}
\caption{{\bf Upper:} Transit depths of V1298 Tau e as a function of wavelength, reduced with \texttt{Eureka!}. The blue and red curves show the best-fit transmission spectra from our free retrieval with and without CS$_2$, respectively. Including CS$_2$ yields a decisive improvement in the Bayesian evidence ($\Delta \ln Z = 33.2$), corresponding to an 8.4$\sigma$ preference. {\bf Lower:} Absorption cross sections of the nine dominant molecular species included in our free retrieval and forward model. While a wider range of additional species were explored using \texttt{POSEIDON} opacity database, CS$_2$ uniquely reproduces both the $\sim$4.6--4.7~$\mu$m feature and the sharpened peak near $\sim$4.3~$\mu$m that cannot be explained by CO$_2$ or CO alone.}\label{fig:spectrum}
\end{figure}

We observed one transit of V1298 Tau e with the James Webb Space Telescope (JWST, \cite{Gardner}) using its NIRSpec/G395H mode. We reduced the data with both \texttt{Eureka!} \cite{eureka} and \texttt{exoTEDRF} \cite{Radica2024} and found consistent wavelength-dependent light curves. After correcting for instrumental systematics and modeling spot-crossing events,
we obtained a transmission spectrum spanning 2.7--5.2~$\mu$m (Fig.~\ref{fig:spectrum}). We first analyzed the spectrum using a free retrieval with \texttt{POSEIDON} \cite{POSEIDON}. We allowed the abundances of a broad suite of molecules to vary independently. We detected CO$_2$, CS$_2$, and CH$_4$ at high significance (5.8$\sigma$ or higher); we found tentative evidence for HCN, CS, and CO (1.9--3.6$\sigma$); and no evidence for H$_2$O, OCS, SO, SO$_2$, or H$_2$S (see Tab. \ref{tab:retrieval} for volume mixing ratios or VMR). We cross-checked our \texttt{POSEIDON} analysis with an independent free retrieval using the \texttt{petitRADTRANS} code \cite{Molliere2019}. 

Of particular interest is our 8.4$\sigma$ detection of CS$_2$ corresponding to a log(VMR$_{\rm CS_2}$)$=-3.90^{+0.56}_{-0.74}$.  In a hydrogen-rich atmosphere, CS$_2$ should be negligible since H$_2$S is by far the favored sulfur-bearing species under equilibrium chemistry \cite{easychem,Benneke}. However, V1298~Tau is a young, active star with $\sim$100$\times$ the UV output than the present-day Sun \cite{Duvvuri}. Intense UV irradiation can photolyze CH$_4$ and H$_2$S and initiate radical-driven carbon-sulfur chemistry that ultimately forms CS$_2$ (\cite{Veillet}, see Sec.~\ref{sec:forward_model}).

Our detection of CS$_2$ is based on two prominent features in the transmission spectrum (Fig.~\ref{fig:spectrum}): an excess absorption feature at 4.6--4.7~$\mu$m that is narrower than the CO band, and a sharp peak near 4.3~$\mu$m that cannot be reproduced by CO$_2$ by itself. We tested a broad range of additional molecular species available in the \texttt{POSEIDON} opacity database%
\footnote{\url{https://poseidon-retrievals.readthedocs.io/en/latest/content/opacity_database.html}}
and found that no other molecule could fit the data. Formally, including CS$_2$ improved the best-fit reduced $\chi_{\rm red}^2$ from 1.77 to 1.08 for 99 degrees of freedom and increased the natural logarithm of the Bayesian evidence by $\Delta \ln Z = 33.2$. Taken together, the multiple spectral signatures, the decisive gain in Bayesian evidence, and the fact that our photochemical forward models (described below) successfully reproduced the observed CS$_2$ VMR support the presence of CS$_2$ on V1298 Tau e.

We performed a suite of one-dimensional, radiative-convective-photochemical forward models for V1298 Tau e. We computed pressure-temperature profiles with \texttt{PICASO} \cite{PICASO}, chemical abundances with \texttt{Photochem} \cite{Photochem}, and transmission spectra with \texttt{POSEIDON}. Our model grid spanned atmospheric metallicity%
\footnote{Here, [M/H] = $\log_{10}(n_M/n_H)$ - $\log_{10}(n_M/n_H)_\odot$, where $n_M$ is the number density of atoms heavier than helium, $n_H$ is the number density of hydrogen atoms, and $\odot$ denotes the solar value.}
[M/H], carbon-to-oxygen ratio C/O, internal temperature $T_{\rm int}$ (a measure of the internal heat flux of the planet), vertical eddy diffusion coefficient $K_{\rm zz}$ (the strength of vertical mixing), and sulfur enrichment [S/M] \footnote{[S/M] denotes the sulfur abundance relative to the other heavy elements, in dex}. The best-fitting models provide satisfactory fits to the data ($\chi_{\rm red}^2 \approx 1.2$, Fig. \ref{fig:forward_model}), and reproduced the inferred CS$_2$ abundance of $\log(\mathrm{VMR}_{\rm CS_2}) \approx -4$ (Fig. \ref{fig:vmr_cs2}). This demonstrates that the detected CS$_2$ is physically plausible in the atmosphere of V1298 Tau e.

Our forward models favor a relatively low metallicity [M/H] $=0.63\pm0.24$, a broadly super-solar carbon-to-oxygen ratio C/O $=0.69\pm0.18$ (the solar values is close to 0.55 \cite{Asplund09}), a modest eddy diffusion coefficient of $\log(K_{\rm zz}/{\rm cm^2\,s^{-1}})=6.1\pm0.6$, and an internal temperature of $T_{\rm int}=303\pm72$ K. These conditions allow CH$_4$ to remain sufficiently abundant to seed carbon--sulfur chemistry while still lofting CS$_2$ into the pressures probed by transmission spectroscopy. Strong interior heating tends to suppress CH$_4$ \cite{Morley,Sing,Welbanks,Yu}, driving the atmosphere away from the CH$_4$-rich regime required to produce CS$_2$. The non-detections of H$_2$O and OCS, together with the detection of CS$_2$, are suggestive of a relatively high C/O atmosphere. The absence of SO$_2$ at these cool temperatures is also consistent with photochemical atmosphere grids that have explored a range of planetary parameters \cite{Mukherjee,crossfield2025}. We acknowledge several caveats to the atmospheric parameters presented above. First, our observations span 2.7--5.2~$\mu$m, a wavelength range with limited sensitivity to key species such as H$_2$O. Second, the IR opacity of CS$_2$, particularly at high temperatures and low pressures, remains poorly constrained \cite{Gordon,Huang}. Finally, sulfur reaction networks, especially the oxidation of CS$_2$ in H/He-dominated atmospheres is uncertain \cite{Veillet}. 

The next planet interior to V1298 Tau e, planet b, has also been observed with JWST in the same mode \cite{Barat}. Planet b has a period of 24 days, a size of  9 $R_\oplus$ and a mass of $\sim$13 $M_\oplus$. The two planets are siblings in the same system, with broadly similar bulk properties, and both likely represent young precursors of the mature sub-Neptune population. Both planet b and e have low atmospheric metallicities ([M/H] $=0.6^{+0.4}_{-0.6}$ and $0.63\pm0.24$, respectively) compared to mature sub-Neptunes (where [M/H] is typically  $\gtrsim 2$ \cite{Madhusudhan,Beatty,Benneke}). This suggests that, as planets age, atmospheric escape preferentially removes H/He and increases [M/H] \cite{Owen2017,Ginzburg}. 

On planet b, H$_2$O and CO$_2$ were securely detected, while SO$_2$ rather than CS$_2$ was the dominant sulfur-bearing photochemical product. A previous analysis interpreted these results as evidence of strong vertical mixing and a high internal temperature of $\sim$500 K, possibly heated by obliquity tides \cite{Barat,Millholland2019}. A hotter interior would suppress CH$_4$ and favor oxidized sulfur species such as SO$_2$ \cite{Mukherjee}. By contrast, $T_{\rm int}$ of planet e is largely consistent with planetary evolutionary models without tidal heating \cite{Tejada}. In other words, obliquity tides probably do not operate on planet e, as they require chance capture into a Cassini state \cite{MillhollandNatAs}. If so, CH$_4$ can survive more readily, enabling the reduced carbon-sulfur photochemistry that produces CS$_2$ on planet e. The comparison between V1298 Tau b and e is especially powerful because it isolates atmospheric physics within a common stellar environment: even within the same system, differences in internal heat, vertical transport, and bulk composition can produce qualitatively different photochemical end products.

Yet another possibility is that V1298 Tau b and e are chemically distinct because they accreted gas from different regions of the protoplanetary disk \cite{Oberg,Bergin,Walsh}. Our analysis tentatively favors a super-solar C/O ratio for planet e (C/O $=0.69\pm0.18$), while planet b has a lower value (C/O $=0.22^{+0.06}_{-0.05}$ \cite{Barat}). The planets in the V1298 Tau system are all near mean-motion resonances \cite{David2019,Livingston}, an orbital architecture strongly suggestive of prior inward disk migration \cite{Kley,Mills,Izidoro}. Planets that experienced long-range migration may have formed further out in the disk where chemical environment is different. It is possible that planet e formed exterior to the water snowline, while planet b formed interior to it. Beyond the water snowline, condensation of water ice depletes oxygen from the gas phase and raises the C/O ratio of the accreted gaseous envelope, perhaps explaining the compositional difference of these now neighboring planets. Future observations over a broader wavelength range will be needed to better constrain the water abundance and C/O ratio of V1298 Tau e to enable a more definitive test of this scenario.

\bibliography{scibib}

\bibliographystyle{Science}



\section*{Acknowledgments}

We thank Michael Zhang, Yubo Su, 
Roberto Tejada Arevalo, Saugata Barat, Andrew Vanderburg, Xianyu Tan and Michael Liu for helpful conversations. As we prepared our manuscript, we became aware of the work by Triantafillides et al. \cite{triantafillides2026} who also found evidence of CS$_2$ on WASP-80 b.\\
\textbf{Funding:} This work was supported by JWST GO Program 5882.

\noindent \textbf{Author Contributions:} F.D. led the analyses and writing of this manuscript. E.P. conceived the project and contributed to analyses. J.L. led the \texttt{exoTEDRF} data reduction and \texttt{petitRADTRANS} retrieval. N.W. and S.M. contributed to the photochemical modeling. Z.H. contributed to the JWST data analyses and free retrieval. I.C. and J.O. contributed to the initial proposal and the interpretation of the results. K.M. contributed to the dynamical modeling of the system.

\noindent\textbf{Competing interests:} The authors declare no competing interests.

\noindent\textbf{Data and materials availability:} Data used in this manuscript is archived on MAST. Additional data products can be obtained from F.D. upon reasonable request. Data used in this project was obtained under JWST GTO program 5882 (PI Dai \& Petigura) and can be downloaded from the Mikulski Archive for Space Telescopes (MAST; https://mast.stsci.edu). This study makes use of the following publicly-available packages: \texttt{Eureka!}, \texttt{Photochem}, \texttt{PICASO}, \texttt{POSEIDON}, \texttt{VULCAN}, \texttt{MultiNest}, \texttt{exoTEDRF}, and \texttt{petitRADTRANS}.

\baselineskip24pt

\section*{Supplementary materials}

\renewcommand{\thesection}{S\arabic{section}}
\setcounter{section}{0}
\section{NIRSpec Observations} \label{sec:nirspec}

We observed V1298 Tau with the Near Infrared Spectrograph (NIRSpec) on JWST in the Bright Object Time Series (BOTS) mode. The observation was conducted on Barycentric Julian Date BJD = 2460941 (UT 2025 September 22–23) over a total duration of 15.3 hours, covering approximately twice the transit duration of V1298 Tau e. We used the G395H/F290LP grating/filter combination, with spectra dispersed across both the NRS1 and NRS2 detectors through the $1.6^{\prime\prime} \times 1.6^{\prime\prime}$ fixed slit aperture. We employed the SUB2048 subarray and the NRSRAPID readout pattern, resulting in a total of 6,750 integrations containing 8 groups each. Our subsequent analysis began from the raw, uncalibrated (\texttt{*uncal.fits}) files archived at MAST. We excluded the initial segment of the light curve, which contained a partial transit of planet d, from the present analysis.

\section{\texttt{Eureka!} Reduction} \label{sec:eureka}
\begin{figure}[t!]
\begin{center}
\includegraphics[width=0.49\textwidth]{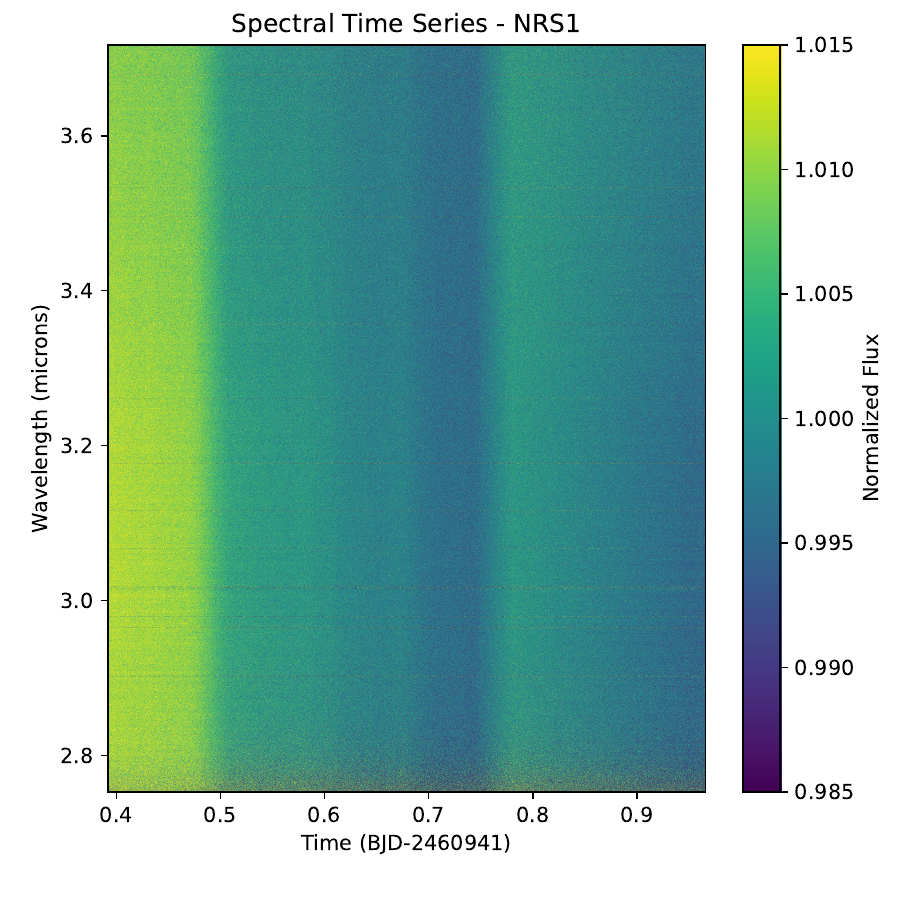}
\includegraphics[width=0.49\textwidth]{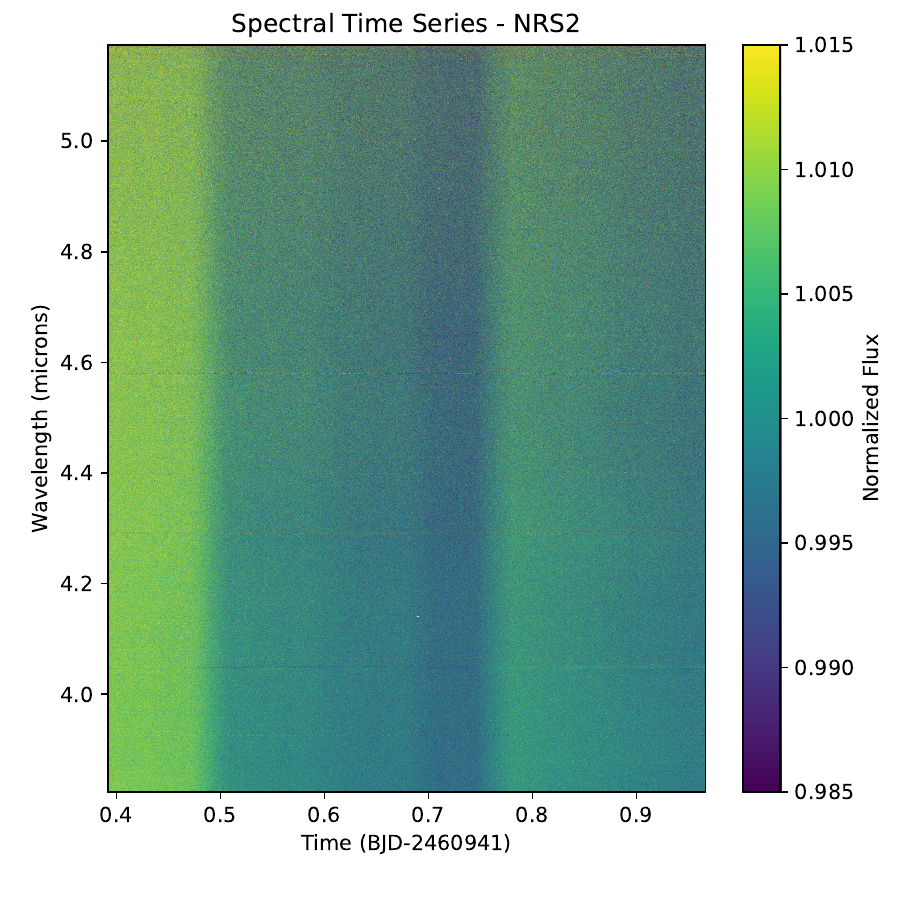}
\end{center}
\caption{ {\bf JWST/NIRSpec observation of a transit of V1298 Tau e.} The left and right panels show the observations in NRS1 and NRS2 respectively. The observations were both reduced with \texttt{Eureka!}.}\label{fig:2d}
\end{figure}

\begin{figure}[t!]
\begin{center}
\includegraphics[width=0.49\textwidth]{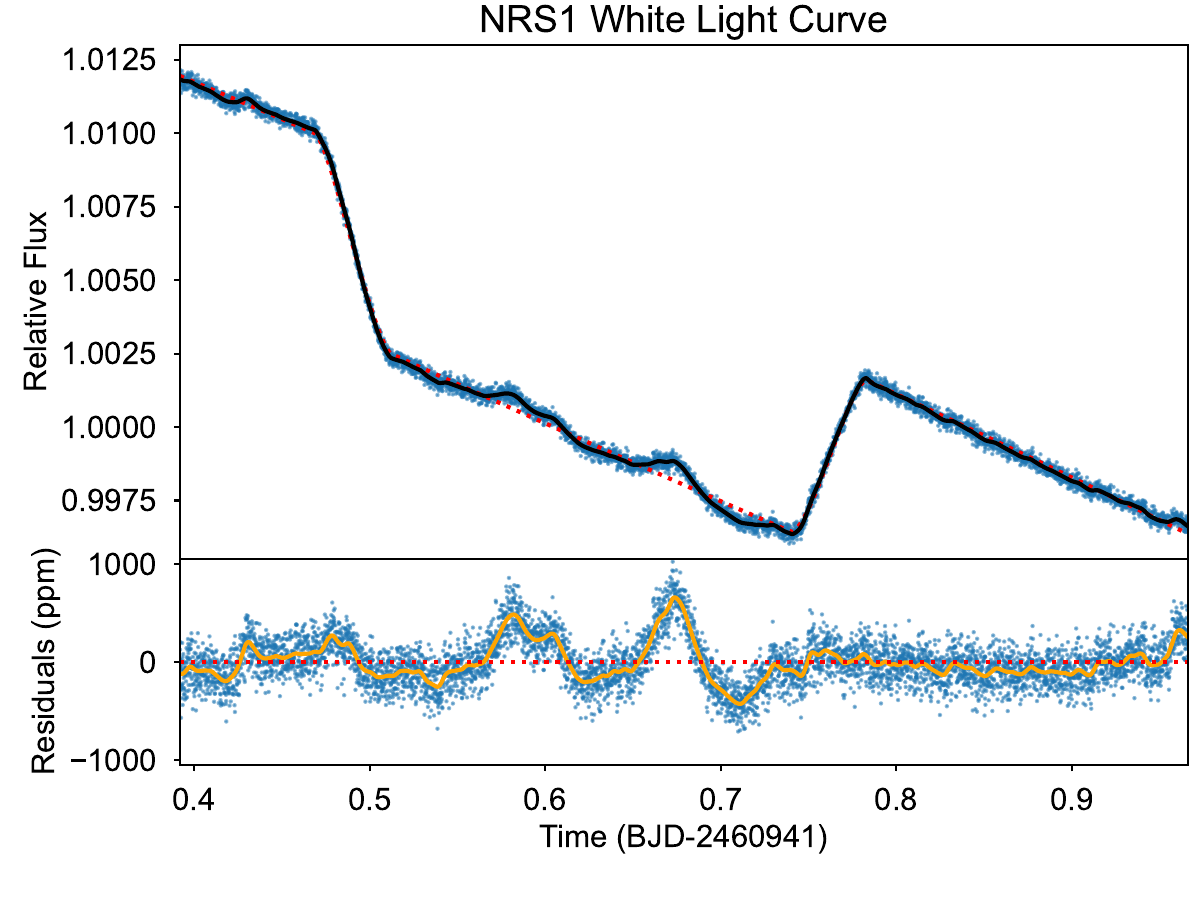}
\includegraphics[width=0.49\textwidth]{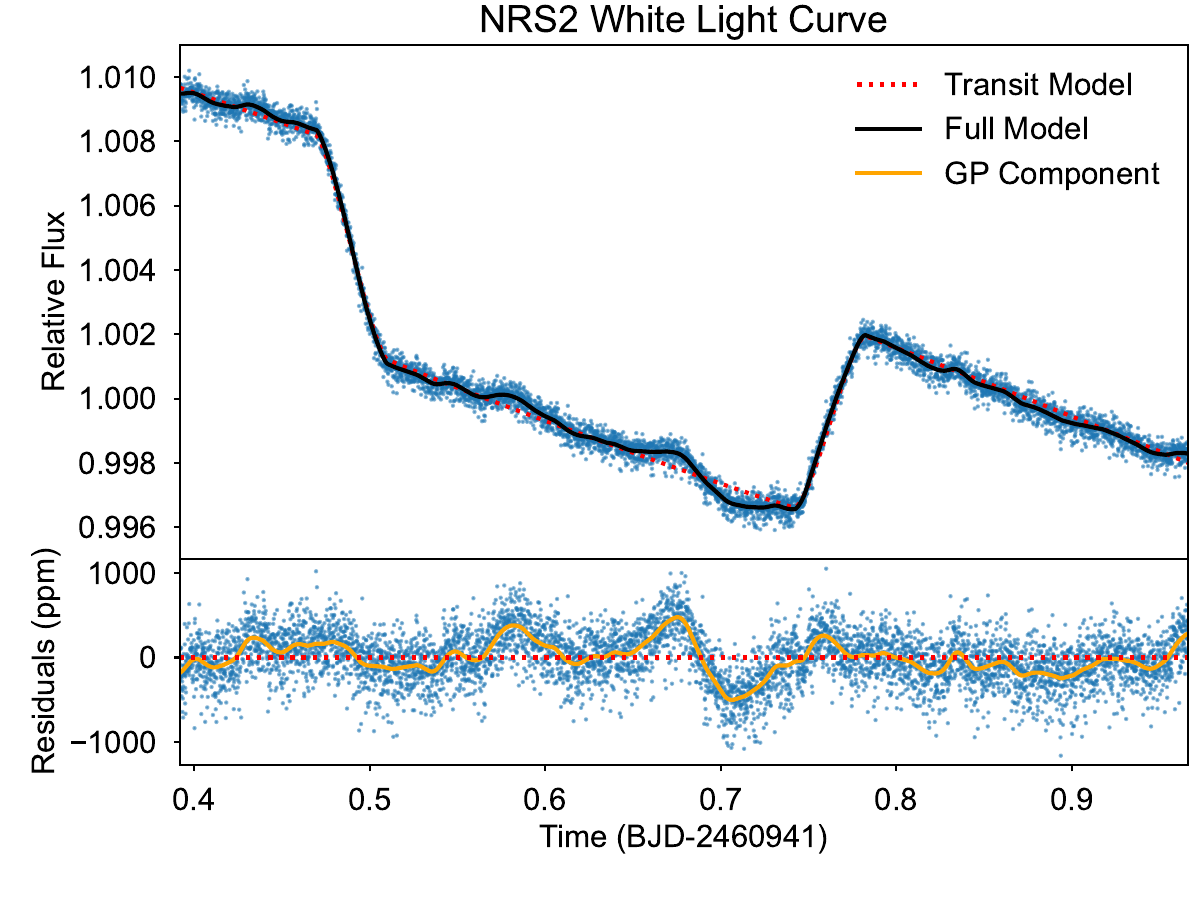}
\end{center}
\caption{{\bf White (wavelength integrated) transit light curves of V1298 Tau e.} Blue symbols shows the data; the red dotted line shows the best-fit \texttt{batman} transit model; the orange curve shows
the \texttt{celerite2} Gaussian Process model for spot-crossing events and other short-term variabilities; the black curves show the full model.}\label{fig:lc}
\end{figure}

We reduced the NIRSpec data using the publicly available \texttt{Eureka!} package (version 1.1 \cite{eureka}). Our reduction made use of the first four \texttt{Eureka!} stages. Stage 1: convert up-the-ramp counts to slopes; Stage 2: perform flat-fielding, remove bad pixels, and correct for 1/$f$ noise at the group level; Stage 3: perform optimal extraction to obtain a time series of 1-D spectra \cite{Horne1986}; Stage 4: produce white and spectroscopic light curves by binning the time series of 1D spectra along the wavelength axis. We followed standard procedures adopted in previous NIRSpec/G395H time-series analyses  \cite{May,Benneke,Barat}. Our \texttt{Eureka!} control and parameter files are available on Zenodo. The spectral time series and corresponding white light curves are shown in Fig.~\ref{fig:2d} and Fig.~\ref{fig:lc}.

We visually inspected light curves generated with each of the 1,426 columns from NRS1 and found 20 that exhibited anomalous behavior associated with stellar spectral lines or high variability; and excluded them from our subsequent analysis. The excluded columns were 221, 312, 313, 315, 316, 336, 337, 388, 389, 391, 393, 394, 538, 629, 752, 836, 911, 1100, 1154, and 1155. We followed the same procedure for NRS2 and removed the following 14 columns: 156, 365, 536, 706, 707, 708, 869, 914, 1105, 1109, 1143, 1970, 2009, and 2010.

\section{Light Curve Modeling}\label{sec:lc}

We began with the extracted light curves from the Stage 4 products of \texttt{Eureka!}. The NRS1 and NRS2 datasets are modeled independently. We first modeled the white transit light curves using \texttt{batman} \cite{batman}. We fixed the orbital period to 48.677714 days found by \cite{Livingston} and held $e$ = 0 since \cite{Livingston} only found an upper limit of 1.24\%. We allowed the following parameters to vary freely within a uniform prior range: time of conjunction $t_c$, planet-to-star radius ratio $R_p/R_\star$, scaled semi-major axis $a/R_\star$, cosine of the orbital inclination $\cos i$, and a linear limb-darkening coefficient $u$. We found that a quadratic limb-darkening law yielded indistinguishable results. We modeled long-timescale ($\gg$ 1 hour) stellar/instrumental variability as a quadratic polynomial with parameters $f_0$, $f_1$, and $f_2$. The prior range of the parameters are presented in Tab. \ref{tab:para}.

Spot-crossing events \cite{Sanchis-Ojeda} are clearly visible during the transit of V1298 Tau e (Fig.~\ref{fig:lc}). We explored two approaches to mitigate their impact on the transmission spectrum: 1) removing the data points between BJD=2460941.56 to 2460941.72 which are most severely affected by spots/faculae; 2) we also modeled spot-crossing events and other short-timescale ($\sim$1 hour) variability with a Gaussian Process (GP) using a Matérn-3/2 kernel implemented in \texttt{celerite2} \cite{celerite1,celerite2}. These two approaches produced nearly identical transmission spectra (Fig.~\ref{fig:spot_crossing}). We therefore adopted the GP model, which provides a flexible description of the correlated variability while retaining nearly all of the data. The GP component is defined as follows:
\begin{equation}
   \Sigma (\tau) = \sigma_{\rm gp}^2 \left( 1+\frac{\sqrt{3}\tau}{\rho}\right) \exp(-\frac{\sqrt{3} \tau}{\rho})
\end{equation}
We adopted log-uniform priors on the GP hyperparameters: $\log(\rho/\mathrm{day})$ for the correlation timescale (fixed across all wavelengths) and $\log(\sigma_{\rm gp})$ for the GP amplitude (wavelength-dependent). We also include a jitter term $\log(\sigma_{\rm jit})$ (wavelength-dependent) to capture any additional white noise term.

Posterior sampling was performed using \texttt{emcee} \cite{emcee} with 128 walkers and 50,000 steps. We discarded the first 5,000 steps as burn-in. Convergence is verified by ensuring that the chain lengths exceed 100 times the various autocorrelation timescales. The posterior distributions are summarized in Tab. ~\ref{tab:para}.

Following the white-light analysis, we modeled spectroscopic light curves by binning the G395H spectra using every 30 neighboring pixel resulting in a total of 114 spectroscopic light curves. Each is modeled using the same framework as above, except that we fix $t_c$, $a/R_\star$, cos$i$, and the GP timescale $\log(\rho)$ at the best-fit values obtained from the white-light curve analysis. The remaining transit parameters were allowed to float with the same prior ranges. Posterior sampling also followed the same \texttt{emcee} setup. The resulting transmission spectrum ($R_p/R_\star$ or transit depth versus wavelength) is shown in Fig. 1.

\begin{table}[t!]
\centering
\begin{tabular}{llll}
\textbf{Stellar Parameters}                                         & \textbf{Value}                        & \textbf{Source}                \\ \hline
Effective Temperature ($T_{\rm eff}/$K)                                           & $4970 \pm 120$                       & \cite{David2019}        \\     
Stellar Mass ($M_\star/M_\odot$)                                                    & $1.10 \pm0.05$            & \cite{David2019}         \\ Stellar  Radius ($R_\star/R_\odot$)                                                  & $1.305 \pm 0.070$         & \cite{David2019}                           \\ 
Age ($\tau/$ Myr)                                                  & 23$\pm$4                                    & \cite{David2019}      \\ \hline 
\textbf{White Light Curve Parameters}                                          & \textbf{Prior}       & \textbf{Posterior NRS1}  & \textbf{Posterior NRS2}                                 \\ \hline
Orbital Period ($P_{\rm orb}$/days)                                               & fixed at 48.677714                        \\  
Eccentricity   ($e$)                                                     & fixed at 0 \\  
GP Amplitude ($\log \sigma_{\rm gp}$)   & [-10,0] & $2.84 \pm 0.15$ &  $2.11 \pm 0.22$ \\
GP Timescale ($\log (\rho$/day))  & [-5,0] & $-4.29 \pm 0.16$ &  $-3.89 \pm 0.21$ \\
Flux Jitter ($\log \sigma_{\rm jit}$)   & [-15,-1] & $2.27 \pm 0.15$ & $1.98 \pm 0.12$ \\
Limb Darkening Coefficient (u)  & [0,1]  &$0.083 \pm 0.042$& $0.067 \pm 0.065$\\
Time of conjunction ($t_c$/$\mathrm{BJD}-2460941)$                         & [0.4,0.8]    & $0.6260 \pm 0.00046$&      $0.6259 \pm 0.00058$             \\  
Planet-to-Star Radius Ratio ($Rp/R_\star$) & [0,0.2] & $0.0800 \pm 0.0010$&$0.07937 \pm 0.0013$  \\
Scale Semimajor Axis ($a/R_\oplus$) & [0,100] & $40.54 \pm 0.58$ & $41.47 \pm 0.71$\\
Cosine of Orbital Inclination (cos$i$) & [0,0.2] & $0.0173 \pm 0.00050$ & $0.0164 \pm 0.00066$\\
Quadratic Flux Trend ($f_o$) & [0, 10$^5$] & 83978$\pm$ 11 & 34887$\pm$7 \\
Quadratic Flux Trend ($f_1$/~day$^{-1}$) & [-10$^4$, 10$^4$] & -2080$\pm$ 130  & -643$\pm$ 66 \\
Quadratic Flux Trend ($f_2$/~day$^{-2}$) & [-10$^4$, 10$^4$] & -310$\pm$ 240 & -100$\pm$ 120\\\hline 
\end{tabular}
\caption{Stellar and Planetary Parameters of V1298 Tau e}\label{tab:para}
\end{table}

\section{\texttt{POSEIDON} Free Retrieval}\label{sec:retrieval}
We carried out a free retrieval analysis using \texttt{POSEIDON} \cite{POSEIDON}, which solves radiative transfer in a one-dimensional, plane-parallel atmosphere under hydrostatic equilibrium in transmission geometry. The atmosphere is discretized into 100 layers spanning pressures from $10^{-7}$ to 100 bar. The model includes opacity contributions from key molecular absorbers expected on a planet like V1298 Tau e under both equilibrium chemistry \cite{easychem} and photochemistry \cite{Mukherjee}: H$_2$O, CO, CO$_2$, CH$_4$, NH$_3$, H$_2$S, SO, SO$_2$, HCN, OCS, CS$_2$, and CS. Molecular abundances parameterized by volume mixing ratio (VMR) are assumed to be vertically constant and are treated as free parameters. We fixed various stellar and planetary properties at their median values reported in \cite{Livingston} and reproduced in Tab. \ref{tab:para}.

We explored both isothermal and non-isothermal pressure--temperature (P--T) profiles, with the latter parameterized following \cite{Madhusudhan2009}. The non-isothermal model was not favored by the Bayesian evidence and yielded atmospheric constraints that were qualitatively consistent with those from the isothermal model. We report the isothermal model here. We also tested three cloud prescriptions implemented in \texttt{POSEIDON}: (1) a global gray cloud deck with cloud-top pressure $P_{\rm cloud}$; (2) a \texttt{patchy cloud} model, in which an additional parameter was included to specify the fractional coverage of clear and cloudy regions; and (3) a \texttt{cloud+haze} model, in which the haze opacity is described by a power law with two parameters: a Rayleigh enhancement factor and a power-law slope. Among these models, the simple gray cloud deck yielded the highest Bayesian evidence, and we therefore adopted this cloud prescription in our retrieval. A similar cloud treatment was used by \cite{Barat} in their analysis of V1298 Tau b, facilitating a direct comparison between the two planets.

Additional parameters of the retrieval include the reference radius at 1 bar pressure $R_{\rm p, ref}$ as well as an offset between the NRS1 and NRS2 detectors \cite{Alderson,Sarkar}. High-resolution model spectra ($R \approx 100$k) were computed and subsequently binned to the resolution of the observations. Posterior sampling was performed using the nested sampling algorithm \texttt{MultiNest} \cite{Feroz2009} implemented in \texttt{PyMultiNest} \cite{Buchner2016} with 2000 live points and default settings. Detection significances for individual species were estimated by comparing the Bayesian evidence of the full model with models in which the opacity of that particular species is set to 0.

We report the posterior distribution of our isothermal model in Tab. \ref{tab:retrieval} and Fig. \ref{fig:retrieval}. We detected CO$_2$ (9.6$\sigma$), CS$_2$ (8.4 $\sigma$), and CH$_4$ (5.8$\sigma$) at high significance ($>5\sigma$). We found tentative evidences for HCN (3.6$\sigma$), CS (2.2$\sigma$), and CO (1.9$\sigma$), and non-detection for NH$_3$, OCS, SO, SO$_2$, H$_2$O, and H$_2$S. We caution that non-detection does not rule out the presence of these molecules on the planet. More likely, the limited wavelength coverage of NIRSpec/G395H restricts our sensitivity to these molecules. We encourage future observations in other wavelength ranges. We also emphasize that the conversion between Bayes factor and $\sigma$ significance is optimistically biased and should be treated with caution for marginal detections \cite{Thorngren,Kipping}. We therefore report both the Bayes factor and $\sigma$ significance in Tab. \ref{tab:retrieval}.

 We further explored a broader suite of molecular species available in the \texttt{POSEIDON} opacity database, but found that no alternative species is favored over CS$_2$. CS$_2$ uniquely reproduces both the $\sim$4.6--4.7~$\mu$m feature and the sharpened peak near $\sim$4.3~$\mu$m, which cannot be explained by CO$_2$ or CO alone (Fig.~1). Including CS$_2$ in our model yielded a best-fit model with 98 degrees of freedom and $\chi^2= 106.2$ (reduced $\chi_{\rm red}^2$=1.08); whereas without CS$_2$ there was 99 degrees of freedom, $\chi^2$ increased substantially to $174.9$ (reduced $\chi_{\rm red}^2$=1.77). CS$_2$ is a photochemical product \cite{Beatty}; and the presence of active photochemistry implies that VMRs cannot be straightforwardly mapped onto bulk atmospheric properties such as [M/H] or C/O. We therefore defer constraints on these quantities to our photochemical forward models in the next section.

Unocculted spots and faculae on a young, magnetically active star such as V1298 Tau can imprint wavelength-dependent signals on the measured transmission spectrum through the so-called Transit Light Source effect \cite{Rackham}. V1298 Tau exhibits $\sim 3\%$ optical photometric variability \cite{David2019}, and the spot-crossing events in our light curves provide direct evidence for surface magnetic activity. We therefore tested two stellar-contamination prescriptions implemented in \texttt{POSEIDON}. In the \texttt{one\_spot} model, the stellar disk is represented by a photosphere and a cooler spotted component with a spot-covering fraction. In the \texttt{two\_spot} model, an additional hotter faculae component is included, also with a faculae-covering fraction. We repeated the retrieval analysis using both prescriptions and found that the inferred molecular abundances are fully consistent with those from the fiducial retrieval. In particular, the retrieved CS$_2$ abundance is $\log_{10}({\rm VMR})=-3.90^{+0.56}_{-0.74}$ in the fiducial model, $-3.74^{+0.58}_{-0.71}$ in the \texttt{one\_spot} model, and $-3.91^{+0.57}_{-0.77}$ in the \texttt{two\_spot} model. The limited impact of transit light-source effect is expected because our transmission observations are only in the infrared, where the contrast between active regions and the quiet photosphere is reduced. Moreover, our CS$_2$ detection is driven primarily by localized spectral structure rather than by a broadband slope. We adopted the fiducial retrieval which has the highest Bayesian evidence thanks to its simplicity.

\begin{table}[t!]
\centering
\begin{tabular}{llllc}                      
\textbf{Parameters}                                         & \textbf{Prior}       & \textbf{Posterior} & \textbf{Bayes Factor}   & \textbf{Significance}                      \\ \hline
Retrieved Temperature ($T_{\rm ret}$/K) & [200,1000]& 424$\pm$33 &  \\
Reference Radius at 1bar ($R_{\rm ref}$/$R_{\rm Jup}$) & [0.5,1.2]& 0.90$\pm$0.01 & \\
Cloud Deck Pressure log($P_{\rm cloud}$/bar) & [-7,2]& -3.02$^{+0.44}_{-0.29}$ &  \\
NRS1-NRS2 Offset (ppm) & [-1000,1000]& 111$\pm$30 &  \\
log (VMR CO$_2$) & [-12,-0.3]& -3.93$^{+0.71}_{-1.19}$ & $\Delta \ln Z$=43.4& $\le$9.6 $\sigma$ \\
log (VMR CS$_2$) & [-12,-0.3]& -3.90$^{+0.56}_{-0.74}$& $\Delta \ln Z$=33.2 & $\le$8.4 $\sigma$ \\
log (VMR CH$_4$) & [-12,-0.3]& -4.85$^{+0.54}_{-0.76}$& $\Delta \ln Z$=14.7 & $\le$5.8 $\sigma$ \\
log (VMR HCN) & [-12,-0.3]& -4.44$^{+0.74}_{-1.02}$ & $\Delta \ln Z$=4.8& $\le$3.6 $\sigma$ \\
log (VMR CS) & [-12,-0.3]& -3.45$^{+1.15}_{-2.67}$& $\Delta \ln Z$=1.3 & $\le$2.2 $\sigma$ \\
log (VMR CO) & [-12,-0.3]& -3.71$^{+1.46}_{-4.07}$& $\Delta \ln Z$=0.9 & $\le$1.9 $\sigma$ \\
log (VMR NH$_3$) & [-12,-0.3]& -8.20$^{+2.38}_{-2.53}$ & Not Favored & - \\
log (VMR H$_2$O) & [-12,-0.3]& -8.40 $^{+2.59}_{-2.36}$ & Not Favored& - \\
log (VMR OCS) & [-12,-0.3]& -8.92$^{+1.41}_{-1.95}$ & Not Favored& - \\
log (VMR H$_2$S) & [-12,-0.3]& -8.53 $^{+2.34}_{-2.26}$ & Not Favored& - \\
log (VMR SO) & [-12,-0.3]&  -7.81$^{+2.66}_{-2.72}$ & Not Favored & - \\
log (VMR SO$_2$) & [-12,-0.3]& -9.24$^{+1.82}_{-1.79}$ & Not Favored & - \\ 
Cloud Coverage & [0,1]& 0.67$\pm$0.07 & Not Favored & - \\
Spot Coverage & [0,1]& 0.28$\pm$0.13 & Not Favored & - \\
Spot Temperature (K)& [$T_{\rm eff}$-2000,$T_{\rm eff}$]& 4440$\pm$240 & Not Favored & - \\
Faculae Coverage & [0,1]& 0.09$\pm$0.07 & Not Favored & - \\
Faculae Temperature (K)& [$T_{\rm eff}$,$T_{\rm eff}$+500]& 5180$\pm$150 & Not Favored & - \\
\hline 
\end{tabular}
\caption{Posterior Distribution of our \texttt{POSEIDON} free retrieval}\label{tab:retrieval}
\end{table}

\begin{figure}[t!]
\begin{center}
\includegraphics[width=\textwidth]{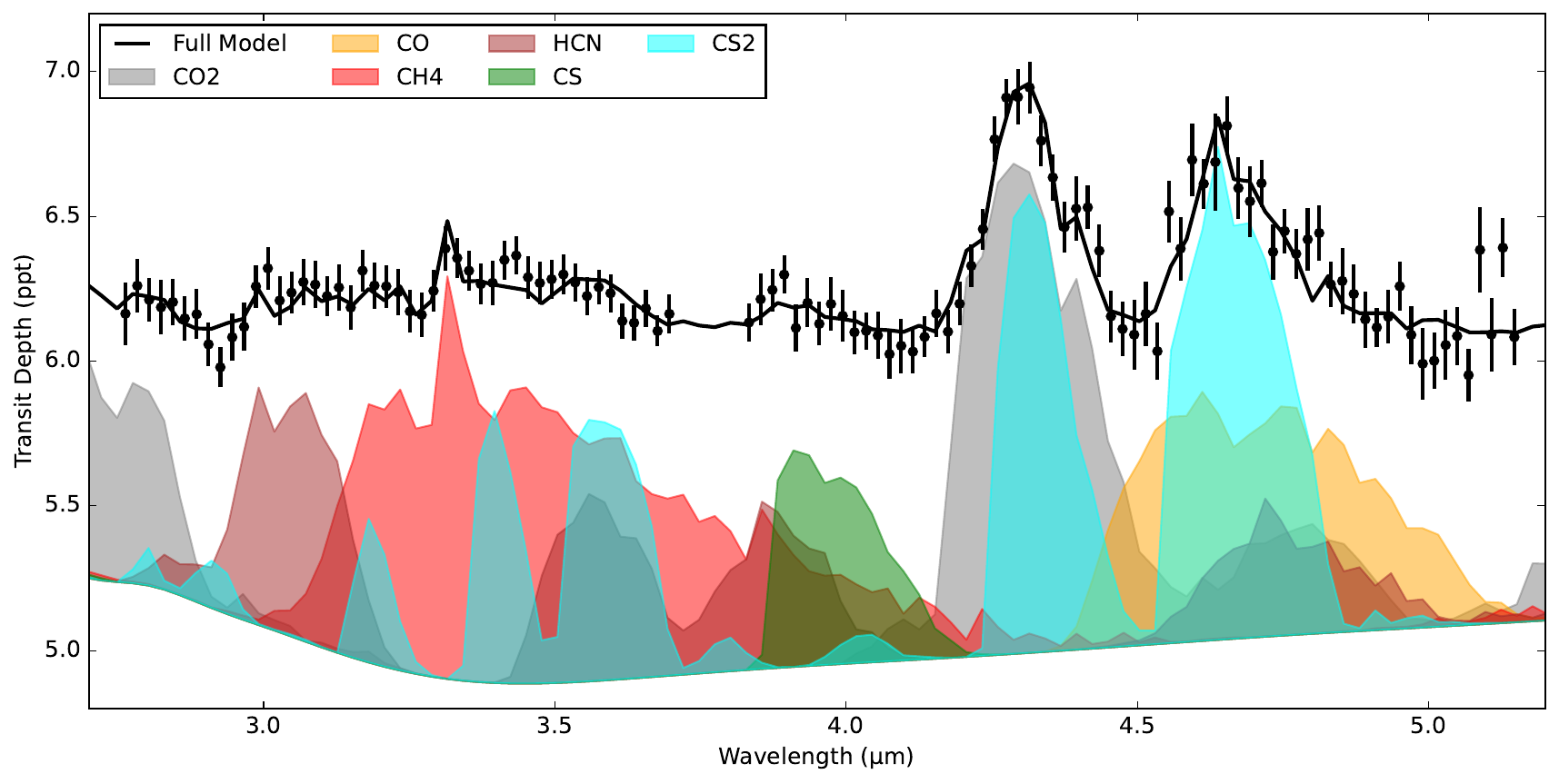}
\end{center}
\caption{{\bf Spectral contribution of detected molecular species} in our \texttt{POSEIDON} free retrieval of V1298 Tau e. Both the observation and the model have been binned to nearest 30 wavelength channels.}\label{fig:spectral_contribution}
\end{figure}

\begin{figure}[t!]
\begin{center}
\includegraphics[width=\textwidth]{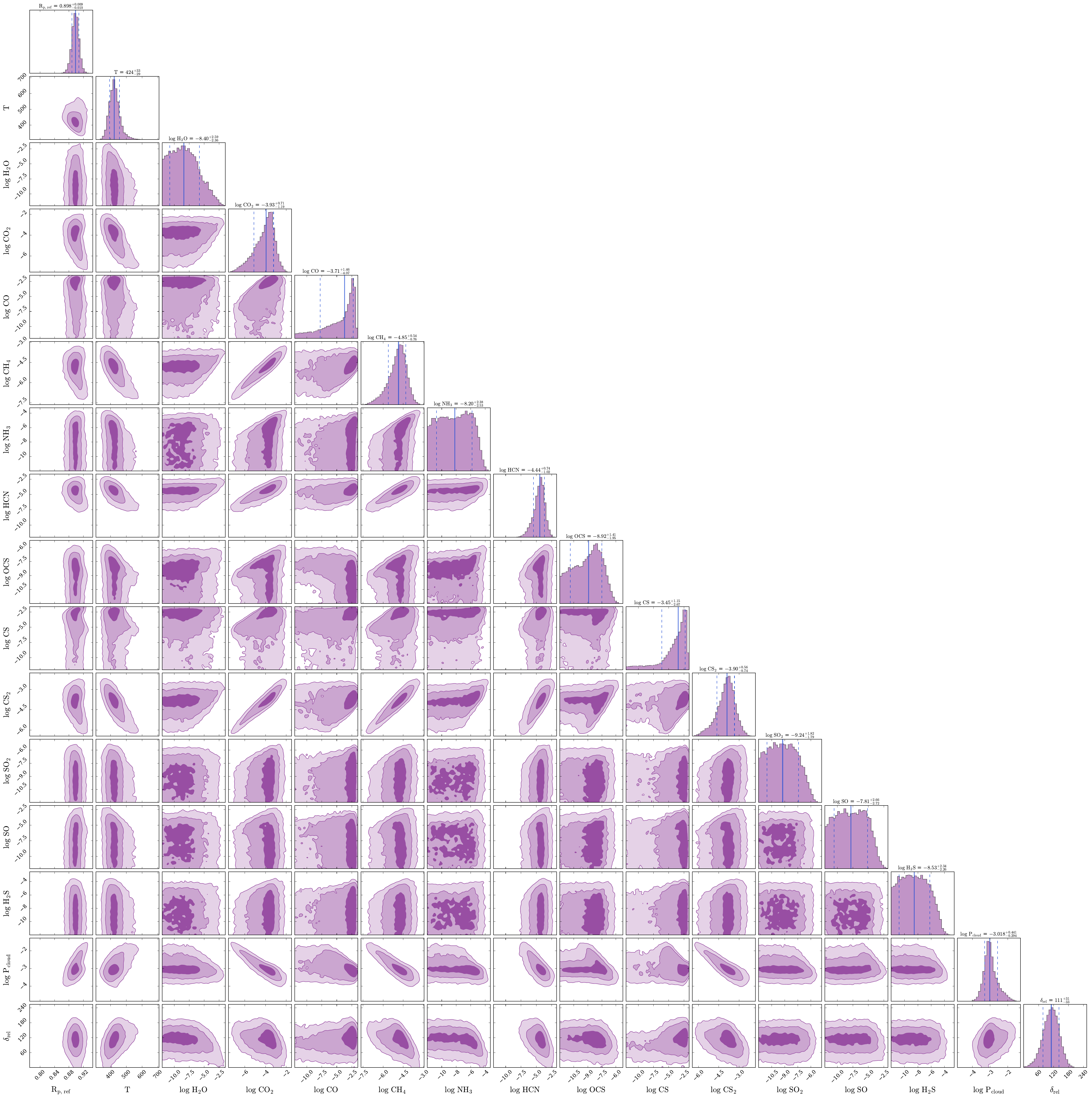}
\end{center}
\caption{{\bf Posterior distributions from our \texttt{POSEIDON} free retrieval of V1298 Tau e.} Results are derived using the nested sampling code \texttt{PyMultiNest}. Full details of the retrieval are given in Sec.~\ref{sec:retrieval}.}\label{fig:retrieval}
\end{figure}

\section{Alternative Analysis with \texttt{exoTEDRF} and \texttt{petitRADTRANS}}
\label{sec:prt}

\begin{figure}[t!]
\begin{center}
\includegraphics[width=\textwidth]{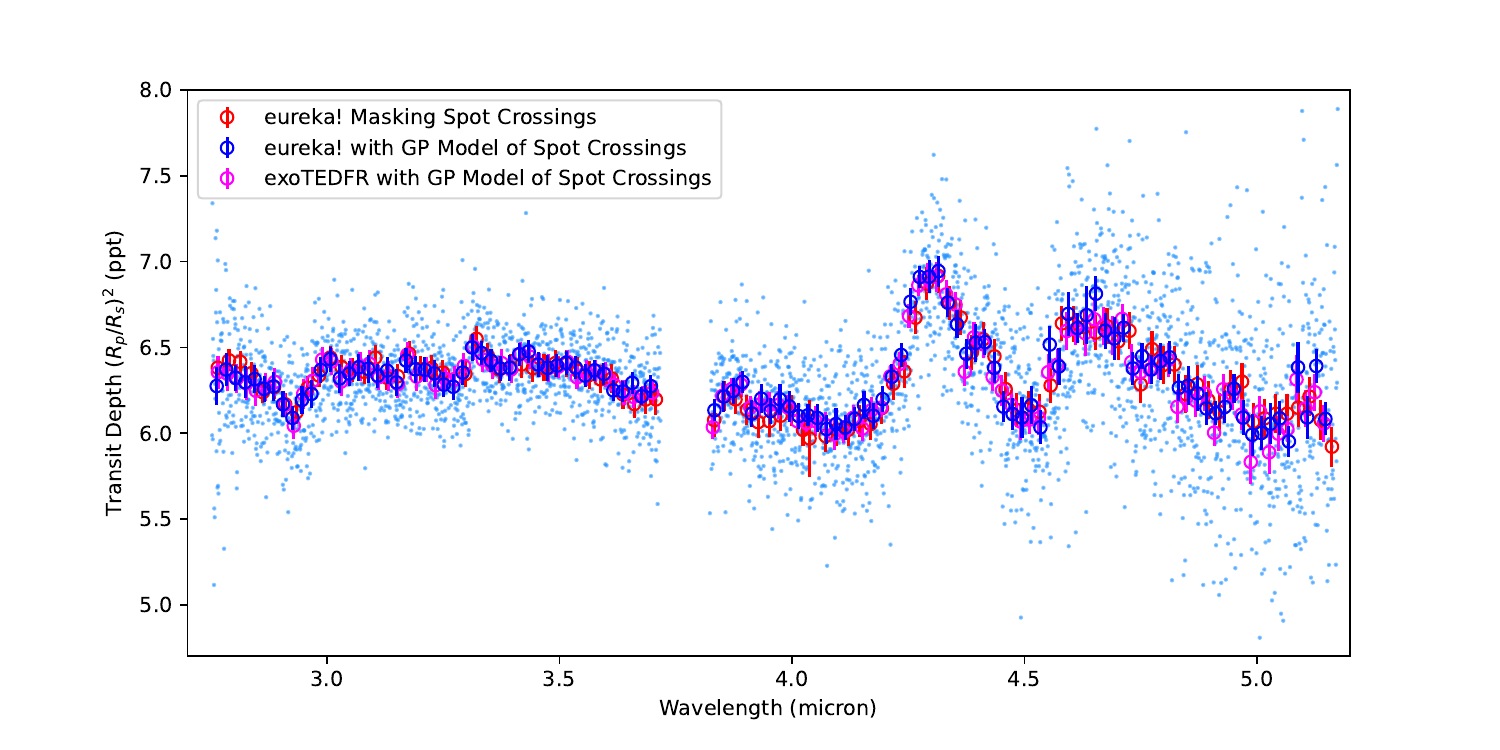}
\end{center}
\caption{{\bf Transmission spectrum of V1298 Tau e extracted using three different methods.} Transit depths are shown after binning over 30 neighboring channels. Red points show the \texttt{Eureka!} reduction in which spot-crossing events were masked, while blue points show the \texttt{Eureka!} reduction in which spot crossings were modeled with a Gaussian process. The masking approach yields transit depths that are overall $\sim3\%$ deeper and has been vertically offset for comparison. Magenta points show the \texttt{exoTEDRF} reduction, also using a Gaussian-process model for spot crossings. All three reductions yield the same spectral features. We adopted the GP-based \texttt{Eureka!} reduction as our fiducial model; blue-filled symbols are the column-by-column transit depths.}
\label{fig:spot_crossing}
\end{figure}

We independently reduced the NIRSpec G395H time series using \texttt{exoTEDRF} v2.4.1 \cite{Radica2024}. Stage~1 applies superbias and dark subtraction, group-level $1/f$ de-striping (median method, 16-pixel NIRSpec trace mask), $7\sigma$ jump detection, ramp fitting, and gain scaling. Stage~2 assigns the WCS, applies wavelength corrections, performs a second $1/f$ pass on masked frames, corrects spatial and temporal bad pixels ($10\sigma$), and removes 10 PCA components. Stage~3 performs box extraction with an 8-pixel half-width per detector. The spectral time series and white light curves are fully consistent with those from \texttt{Eureka!}, see Fig.~\ref{fig:spot_crossing}.

We also performed an independent free retrieval using \texttt{petitRADTRANS} v3 \cite{Molliere2019,Blain2024}. The forward model uses an isothermal 100-layer pressure grid ($\log_{10}P/{\rm bar} \in [-9,+2]$) in transmission geometry, with 12 free-abundance line species (H$_2$O, CO$_2$, CO, CH$_4$, NH$_3$, HCN, OCS, CS, CS$_2$, SO$_2$, SO, H$_2$S), H$_2$/He CIA and Rayleigh scattering, and a solar-ratio H$_2$/He background. Clouds are parameterized as an opaque gray deck (log$\,P_{\rm cloud}$, covering fraction) plus a power-law haze ($\log a$, $\gamma_{\rm scat}$). All log(VMR) priors are uniform on $[-12, 0]$; other priors match Sec.~\ref{sec:retrieval}, with an additional per-NRS2 additive offset $\sim\mathcal{N}(0,10^{-4})$ in $(R_p/R_\star)^2$. Model spectra are convolved to $R=200$. Posterior sampling uses \texttt{MultiNest} \cite{Feroz2009} via \texttt{PyMultiNest} \cite{Buchner2016}. Our \texttt{petitRADTRANS} retrieval also favors the detection of CS$_2$ and yielded VMRs consistent with the \texttt{POSEIDON} retrieval. Our subsequent analysis is based on the data reduction with  \texttt{Eureka!} and retrieval with \texttt{POSEIDON}.

\section{Photochemical Forward Model}\label{sec:forward_model}

To further interpret the observed transmission spectrum, we compare the data to a suite of 1D radiative–convective–photochemical equilibrium (RCPE) models. This approach enables a physical inference of atmospheric properties, accounting for the atmospheric composition, the pressure-temperature (PT) profile, and disequilibrium chemistry.

We first computed radiative–convective equilibrium PT profiles using the \texttt{PICASO} framework \cite{PICASO}. The models span pressures from 100~bar to $10^{-7}$~bar, we adopted the opacities from \cite{Lupu} and \cite{Batalha}. The PT-profile was computed with a median stellar effective temperature and  $a/R_\star$ in Tab. \ref{tab:para} as well as a redistribution parameter {\bf rfacv}$=$0.5 (for full day-night heat redistribution consistent with the low retrieved temperature). We explored a grid in metallicity [M/H]=[0.,0.25,0.5,0.75,1.,1.25,1.5,2.0], C/O = [0.25,0.5,0.75,1.,1.25,1.5] times solar $\equiv0.55$ \cite{Asplund09}, internal temperature $T_{\rm int}$=[50,100,150,200,250,300,350,400,450,500]K, and a vertically constant eddy diffusion coefficient $\log_{10}(K_{zz}/$cm$^2$s$^{-1}$)=[5,6,7,8] resulting a total of 7680 grid points. 

These \texttt{PICASO} PT profiles were fed to the 1D photochemical kinetics model \texttt{Photochem} \cite{Photochem}. We used the S-C-H-N-O chemical network reported in \cite{Tsai2026} after experiment with other reaction networks (discussed shortly). We used the high-energy spectral energy distribution from \cite{Duvvuri} which was constructed from X-ray to UV observations (NICER X-ray telescope, the Space Telescope Imaging Spectrograph and Cosmic Origins Spectrograph instruments on Hubble) combined with empirical models. We also extended the model grid to include super-solar sulfur abundances, [S/M] = 0, 0.5, and 1.0, corresponding to sulfur enhancements of up to 1 dex relative to the other heavy elements.

The resulting PT profiles and volume mixing ratio (VMR) profiles were provided to \texttt{POSEIDON} to generate transmission spectra. While the PT and VMR profiles are held constant, we varied the reference planetary radius at 1 bar, the cloud-top pressure, and the relative offset between NRS1 and NRS2 to obtain the best-fit model. The optimization was done with the \texttt{Nelder-Mead} method in \texttt{scipy.optimize} with default settings. 

\begin{figure}[t!]
\begin{center}
\includegraphics[width=\textwidth]{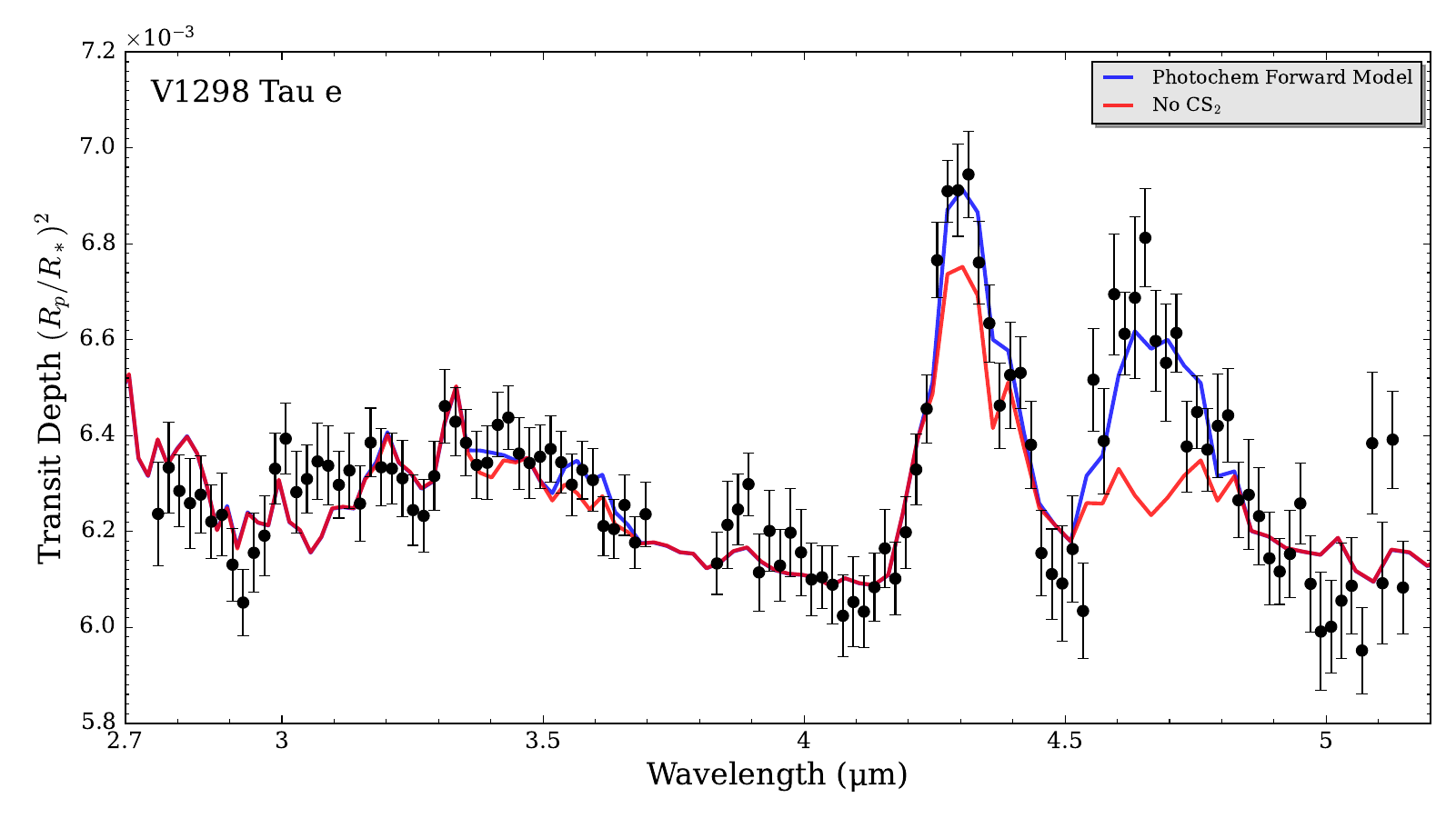}
\end{center}
\caption{{\bf Best-fit \texttt{Photochem} forward model of V1298 Tau e compared to the observed transmission spectrum.} The blue and red curves show models with and without CS$_2$, respectively. The inclusion of CS$_2$ provides a markedly improved fit, capturing spectral features otherwise unexplained.} \label{fig:forward_model}
\end{figure}

\begin{figure}[t!]
\begin{center}
\includegraphics[width=0.8\textwidth]{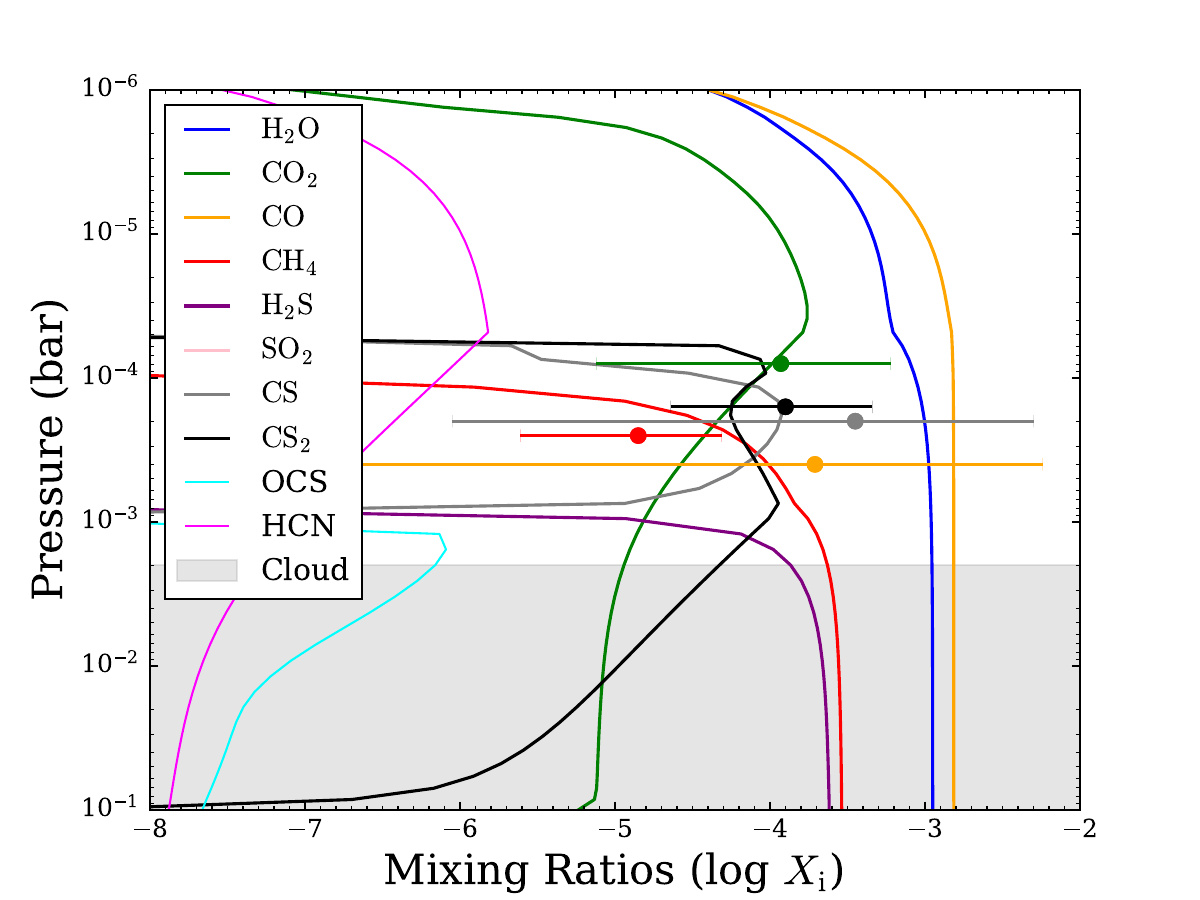}
\end{center}
\caption{{\bf Volume Mixing Ratio (VMR) of the dominant molecular species in V1298 Tau e.} The colored lines are VMR calculated with the kinetic photochemical model \texttt{Photochem}. The errorbars are the VMR of the three species whose detection is favored in our \texttt{POSEIDON} free retrieval. SO$_2$'s VMR is so low that it is off this chart.} \label{fig:vmr}
\end{figure}

\begin{figure}[t!]
\begin{center}
\includegraphics[width=0.8\textwidth]{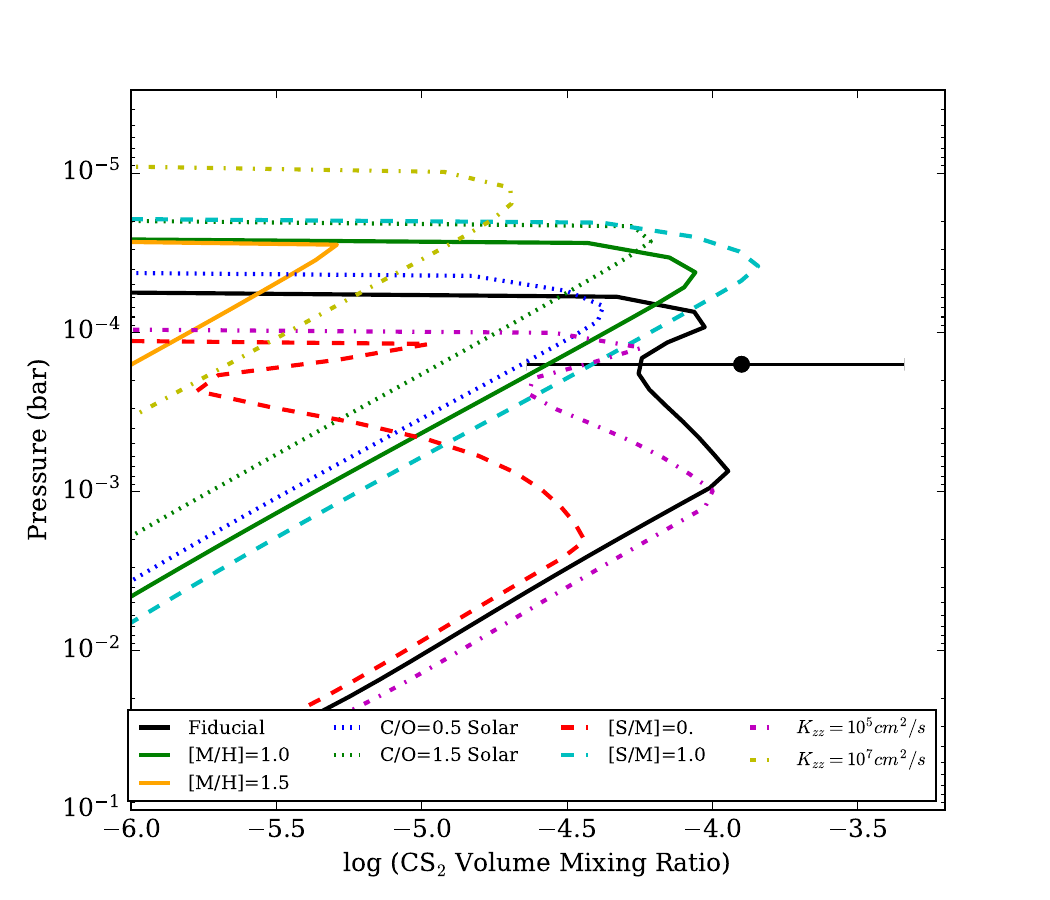}
\end{center}
\caption{{\bf Volume Mixing Ratio (VMR) of CS$_2$ as a function of atmospheric properties.} The fiducial model (black) has metallicity [M/H] = 0.5, carbon-oxygen ratio C/O=1.25 $\times$ solar, $T_{\rm int}=350$K, sulfur abundance [S/M]=0.5, and eddy diffusion constant $K_{\rm zz} = 10^6$cm$^2$ s$^{-1}$. The dependence on each parameter is illustrated by varying one parameter at a time, shown with distinct line styles.} \label{fig:vmr_cs2}
\end{figure}

We evaluated each forward model by computing the $\chi^2$ relative to the observed spectrum and assign probabilities $ \propto \exp(-\chi^2/2)$. Our forward models provide satisfactory fit to the observed spectrum with the best models reaching down to $\chi_{\rm red}^2 \approx 1.2$. For example, the best-fit grid point has a metallicity [M/H] = 0.5, carbon-to-oxygen ratio C/O = 1.25$\times$ solar, internal temperature $T_{\rm int}=350$ K, sulfur abundance [S/M] = 0.5, and eddy diffusion coefficient $K_{\rm zz}=10^6$ cm$^2$ s$^{-1}$ and a $\chi^2=121.3$ (see Fig. \ref{fig:forward_model}). The forward models reproduced the major spectral features across the NIRSpec/G395H wavelength range (see Fig. \ref{fig:forward_model}). Crucially, the forward model was able to reproduce a high VMR of CS$_2$ $\log(\mathrm{VMR}_{\rm CS_2}) \approx -4$ in the pressure region probed by transmission spectroscopy (Fig. \ref{fig:vmr} and \ref{fig:vmr_cs2}).

The $\chi^2$ values across the model grid were used to derive constraints on the atmospheric properties (Fig. \ref{fig:forward_corner}). The forward models favor a relatively low metallicity [M/H] = $0.63\pm0.24$, a super-solar but wide constraint on C/O $=0.69\pm0.18$, a low eddy diffusion log$(K_{\rm zz}/$cm$^2$s$^{-1}$) = 6.1$\pm$0.6, and an internal temperature of $T_{\rm int} = 303\pm72$~K.
The gray cloud deck has a top pressure of $\log(P_{\rm cloud}/{\rm bar}) = -2.77\pm 0.26 $ similar to that found in the free retrieval. Higher-than-solar sulfur abundance is mildly preferred by the models at the $1.9\sigma$ level, although the evidence is not decisive. The fiducial model in Fig. \ref{fig:vmr} and \ref{fig:vmr_cs2} have [S/M] = 0.5. Strong vertical mixing facilitates the lofting of CS$_2$, but preserving CH$_4$ requires sufficiently weak mixing to prevent quenching \cite{Morley,Sing,Welbanks,Yu}; our models favor a relatively low eddy diffusion coefficient of log$(K_{\rm zz}/$cm$^2$s$^{-1}$) = 6.1$\pm0.6$. We note, however, that assuming a vertically constant $K_{\rm zz}$ is a simplification, as CS$_2$ lofting and CH$_4$ quenching likely occur at different pressure levels, from mbar to bar pressures. A high internal temperature may also suppress methane abundance which is the key progenitor for CS$_2$; our model favors an $T_{\rm int} = 303\pm72$~K which is substantially lower than the  $T_{\rm int} \sim 500$~K reported on planet b. The comparatively low atmospheric metallicities inferred for V1298 Tau b and e are consistent with their youth and highly inflated radii: interior-evolution models indicate that maintaining a $\sim 10\,R_\oplus$ radius generally requires a low-metallicity H/He envelope even when the planets were young \cite{Tang2025}.

Under equilibrium chemistry, the minimization of the Gibbs free energy favors H$_2$S, rather than CS$_2$, as the dominant sulfur-bearing species \cite{easychem}. If photochemistry is included, the dominant formation channel of CS$_2$ in a UV-irradiated, H/He-dominated (reducing) atmosphere is as follows:
\begin{enumerate}
    \item Make hydrocarbon radicals from CH$_4$:
    \begin{equation}
       \mathrm{CH_4} \xrightarrow{\mathrm{UV}} \mathrm{CH_3}\xrightarrow{\mathrm{}} \mathrm{CH}
    \end{equation}
    \item Liberate S radical from H$_2$S: 
    \begin{equation}
       \mathrm{H_2S} \xrightarrow{\mathrm{UV}} \mathrm{HS} \xrightarrow{\mathrm{}} \mathrm{S} 
    \end{equation}
    \item Make CS:
    \begin{align}
        \mathrm{CH+S} \rightarrow  \mathrm{CS+H}
    \end{align}
    \item Form CS$_2$:
    \begin{align}
       \mathrm{CS+HS} \rightarrow  \mathrm{CS_2+H} \\
       \mathrm{CS+OCS} \rightarrow  \mathrm{CS_2+CO}
    \end{align}  
\end{enumerate}

Producing CS is usually the bottleneck. However, recently Veillet et al \cite{Veillet} pointed out that including CH$_2$S in intermediate reactions could drastically increase the production of CS and hence CS$_2$. 

    \begin{align}
       \mathrm{CH_4+H} \rightarrow  \mathrm{CH_3+H_2} \\
       \mathrm{CH_3+S} \rightarrow  \mathrm{CH_2S+H}\\
       \mathrm{CH_2S+H} \rightarrow  \mathrm{CHS+H_2}\\
       \mathrm{CHS+S_2} \rightarrow  \mathrm{CS+HS_2}\\
       \mathrm{CS+HS} \rightarrow  \mathrm{CS_2+H}
    \end{align}  

Unfortunately, we could not identify a published opacity list for CH$_2$S, and therefore could not test its presence directly in our retrievals.

\begin{figure}[t!]
\begin{center}
\includegraphics[width=0.8\textwidth]{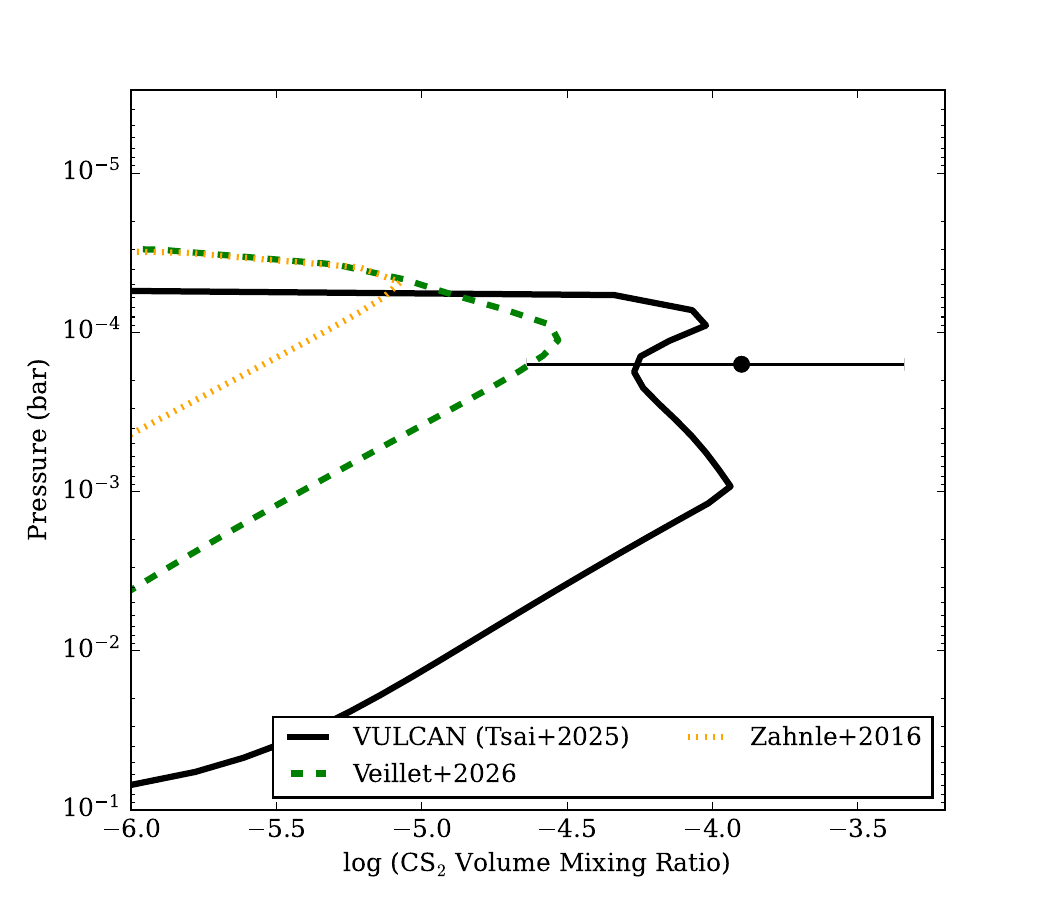}
\end{center}
\caption{{\bf VMR of CS$_2$ predicted by different photochemical reaction networks.} All models are computed using the same atmospheric conditions: metallicity [M/H] = 0.5, carbon-to-oxygen ratio C/O = 1.25$\times$ solar, internal temperature $T_{\rm int}=350$ K, sulfur abundance [S/M] = 0.5, and eddy diffusion coefficient $K_{\rm zz}=10^6$ cm$^2$ s$^{-1}$. The fiducial model (black) uses the recently updated network implemented in \texttt{VULCAN} \cite{Tsai2026}. The dashed green curve shows the result obtained using the network of Veillet et al \cite{Veillet}, while the dotted yellow curve uses the network of Wogan et al \cite{Zahnle2016,Photochem}. The fiducial \texttt{VULCAN} model predicts the highest CS$_2$ abundance, close to the retrieved value shown by the black error bar.}\label{fig:networks}
\end{figure}

\begin{figure}[t!]
\begin{center}
\includegraphics[width=\textwidth]{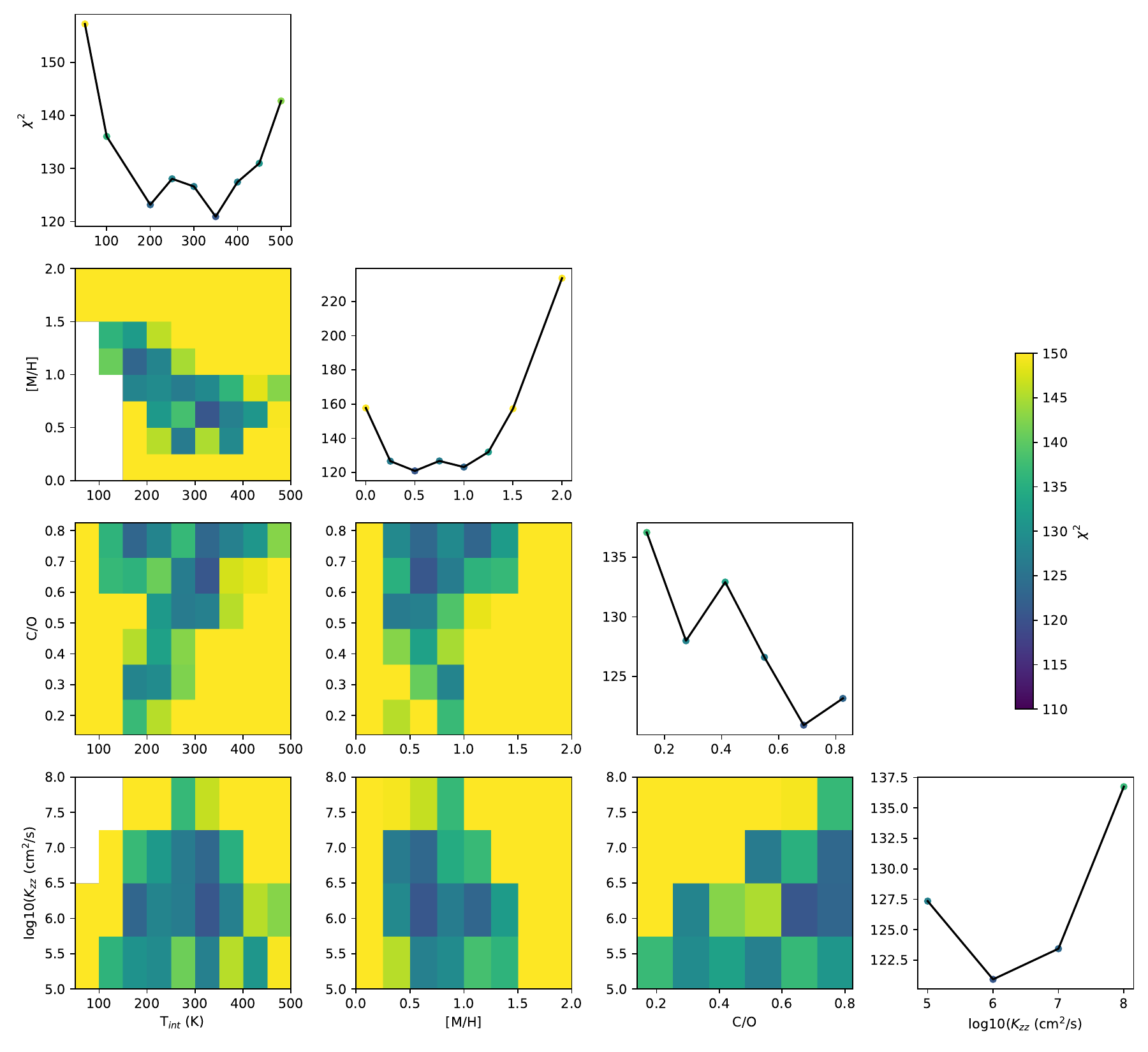}
\end{center}
\caption{{\bf Results of our \texttt{Photochem} forward modeling of V1298 Tau e.} The color coding represents the $\chi^2$ (for 98 degrees of freedom) compared to the observed spectrum. These forward models suggest a low metallicity [M/H] = $0.63\pm0.24$, mildly super-solar C/O $=0.69\pm0.18$, an eddy diffusion log$(K_{\rm zz}/$cm$^2$s$^{-1}$) = 6.1$\pm0.6$, and internal temperature of $T_{\rm int} = 303\pm72$K. Further details are given in Sec.~\ref{sec:forward_model}.} \label{fig:forward_corner}
\end{figure}

We tested three publicly available photochemical networks: 1) Wogan et al. 2025 \cite{Photochem,Zahnle2016}, 2) Tsai et al. 2026 \cite{Tsai2026}, 3) Veillet et al. 2026 \cite{Veillet}. We tested each of these reaction networks under the same atmospheric conditions: metallicity [M/H] = 0.5, carbon-to-oxygen ratio C/O = 1.25$\times$ solar, internal temperature $T_{\rm int}=350$ K, sulfur abundance [S/M] = 0.5, and eddy diffusion coefficient $K_{\rm zz}=10^6$ cm$^2$ s$^{-1}$ i.e. our best-fit forward model. The results are shown in Fig. \ref{fig:networks}.  The network by Wogan et al produced the lowest CS$_2$ (log(VMR)=--5.3), likely because it does not include CH$_2$S, in contrast to the other two networks. The network by Veillet et al. \cite{Veillet} can indeed produce more CS$_2$ ($\log_{10}({\rm VMR})=-4.6$) via their novel CH$_2$S mechanism. The Tsai et al. \cite{Tsai2026} was even more efficient at producing CS$_2$ yielding $\log{10}({\rm VMR})=-4.2$. As an additional consistency check, we ran the Tsai et al. network both within \texttt{Photochem} and natively in \texttt{VULCAN}: their own photochemical model \cite{Tsai2017,Tsai2021}. We obtained nearly identical results. We therefore report photochemical results computed with \texttt{Photochem}, for computational efficiency, using the Tsai et al. network because of its ability to produce CS$_2$ abundantly. 

We point out two important caveats before interpreting our forward models. Firstly, we found that the VMR profiles of various species are broadly consistent with those inferred from the free retrieval (Fig. \ref{fig:vmr}), with the possible exception of water. Our free retrieval could not robustly detect water due to the limited wavelength coverage of NIRSpec/G395H. We encourage follow-up observations to constrain water abundance, which should improve the C/O constraint. Secondly, sulfur photochemistry, particularly pathways involving CS$_2$, remains uncertain \cite{Veillet}. CS$_2$ is a respiratory and neurologically toxic gas due to its strong reactivity with nucleophilic functional groups \cite{Medbery}. Perhaps due to its toxicity, the opacity of CS$_2$ at temperatures higher than room temperature is still missing. The cross-sections in the HITRAN database were measured near room temperatures \cite{Gordon,Huang}. Our analysis therefore relies on an extended grid of CS$_2$ cross sections extrapolated to higher temperatures and lower pressures from available line lists using \texttt{Cthulhu} \cite{Cthulu}. Further theoretical and laboratory work on CS$_2$, especially under H/He-dominated atmospheric conditions, is needed to strengthen its interpretation in exoplanet spectra.

If we take our observed atmospheric constraints at face value, we offer the following interpretation. As pointed out by Mukherjee et al. \cite{Mukherjee}, SO and SO$_2$ are the dominant sulfur photochemical products at high temperatures $>$1000~K and solar C/O ratio. At lower temperatures, CS and CS$_2$ become important S carriers regardless of C/O ratio. This trend is consistent with our strong detection of CS$_2$ in V1298 Tau e and the moderately super-solar C/O ratio inferred from our models, C/O $=0.69\pm0.18$. At still lower temperatures ($\lesssim500$~K), OCS can become an important sulfur-bearing species, in some cases exceeding CS and CS$_2$ as the second-most abundant sulfur carrier after H$_2$S, particularly at low C/O. The simultaneous non-detection of OCS and H$_2$O in our free retrieval may therefore point to a super-solar C/O ratio, which would limit the exposure of CS$_2$ to OH and suppress oxidation pathways such as $\mathrm{CS_2+OH}\rightarrow\mathrm{OCS+HS}$ \cite{Zeng}.

The super-solar C/O ratio inferred for V1298 Tau e contrasts with the sub-solar value reported for V1298 Tau b, C/O $=0.22^{+0.06}_{-0.05}$ \cite{Barat}. Although the two planets have similar retrieved temperatures, $\sim450$~K for V1298 Tau e and $424\pm33$~K for V1298 Tau b, their spectra favor different sulfur species: CS$_2$ in planet e and SO$_2$ in planet b. One possible explanation is that V1298 Tau e, as the outermost planet, formed beyond the disk water snowline and accreted gas with a composition distinct from that of the inner planets, which may have formed interior to the snowline \cite{Oberg}. Atmospheric sulfur abundance may provide an additional diagnostic of formation history, because sulfur is less volatile than C and O and is expected to be largely sequestered in solids in protoplanetary disks \cite{Kama}. Sulfur-bearing atmospheric species may therefore trace the relative importance of pebble versus gas accretion \cite{Crossfield}. However, a recent alternative scenario suggests that sulfur-bearing ammonium salts, such as NH$_4$SH, can enrich the gas disk in sulfur interior to the disk ``salt line'' ($\lesssim3$~AU), potentially imprinting sulfur enrichment on planetary atmospheres without requiring substantial pebble accretion \cite{Ohno}. Our forward models mildly favor super-solar sulfur abundance, [S/M] $\approx0.5$ dex, but only at the $1.9\sigma$ level. Moreover, CS$_2$ oxidation chemistry, especially in H/He-dominated atmospheres, remains insufficiently explored \cite{Veillet} thereby making it difficult to link the observed CS$_2$ VMR to C/O or [S/M] robustly. In short, the implications of the detection of CS$_2$ on the formation history of V1298 Tau e are tantalizing, but not definitive.

\section{Planetary Mass Constraint}\label{sec:mass}

In this work, we adopt the planetary mass measured from transit timing variations (TTVs, \cite{Livingston}). Nevertheless, the amplitude of transmission spectral features provides an independent constraint on the atmospheric scale height, and therefore on the planet’s surface gravity and mass \cite{deWit}:
\begin{equation}
    \Delta D = \frac{2 R_p N H}{R_\star^2},
\end{equation}
where $H$ is the scale height; $N$ is a factor of order unity, typically taken to be $\sim$5 for order-of-magnitude estimates \cite{Seager2000}. Adopting $R_p \sim 10\,R_\oplus$, $R_\star \sim 1.3\,R_\odot$, and a $\sim$900 ppm amplitude for the CO$_2$ feature at 4.3~$\mu$m, we infer a scale height of $\sim$1000 km.

The scale height is related to atmospheric and planetary properties through
\begin{equation}
    H = \frac{k_B T}{\mu m_{\rm pr} g},
\end{equation}
where $k_B$ is the Boltzmann constant, $\mu$ is the mean molecular weight, $m_{\rm pr}$ is the proton mass, and $g$ is the surface gravity,
\begin{equation}
    g = \frac{G M_p}{R_p^2}.
\end{equation}
Combining these relations yields an estimated planetary mass of $\sim 14\,M_\oplus$. We also ran another round of free retrieval with the planetary mass set as a free parameter. We found that planetary mass is degenerate with the reference radius $R_{\rm ref}$. If we impose a stringent prior on $R_{\rm ref}$ based on the transit radius reported by \cite{Livingston}, our retrieval reports a planetary mass of 17.4$\pm2.9 M_\oplus$. In either case, the inferred planetary mass is in good agreement with the TTV-derived value of $15.3 \pm 4.2\,M_\oplus$ \cite{Livingston}. Such a mass places V1298 Tau e as a progenitor for the sub-Neptunes, the most common class of exoplanets in our Galaxy \cite{Fressin,Petigura2013}. We fixed the planetary mass to the median TTV value $15.3M_\oplus$ in our analyses in this work to reduce model complexity.

\section{Tidally Locked?}
At the orbital separation of V1298 Tau e, tidal synchronization may not have occurred within the young age of the system ($\sim20$~Myr). Under the standard equilibrium-tide formalism \cite{Murray,Hui}, the spin-synchronization timescale is given by:
\begin{equation}
\tau_{\rm sync} = \frac{2 \alpha_{\rm p}}{3} \frac{Q_{\rm p}}{k_{2,p}} \frac{M_{\rm p}}{M_\star} \left( \frac{a}{R_{\rm p}}\right)^3 \frac{\Omega_{\rm p, 0}}{n^2},
\end{equation}
where $\alpha_{\rm p}$ is the dimensionless moment of inertia ($I=\alpha_{\rm p} M_{\rm p}R_{\rm p}^2$), $Q_{\rm p}$ is the tidal quality factor, and $k_{2,p}$ is the Love number of the planet. We adopt fiducial values of $\alpha_{\rm p}=0.25$, $Q_{\rm p}=10^5$, and $k_{2,p}=0.3$, comparable to Solar System gas giants \cite{Murray}. We take the initial planetary rotation rate to be 40\% of the break-up frequency, $\Omega_{\rm p,0}=0.4\Omega_{\rm break}=0.4\sqrt{GM_{\rm p}/R_{\rm p}^3}$, motivated by the spin rates of Jupiter and young giant exoplanets \cite{Hsu,Batygin_rotation}. For V1298 Tau e, this gives a synchronization timescale of $\sim300$--$500$~Myr, substantially longer than the system age, suggesting that the planet is unlikely to be tidally locked at present. A non-synchronous, rapidly rotating state would promote efficient day--night heat redistribution, lowering the dayside temperature relative to that of a tidally locked planet that cools primarily from its dayside. This may help explain the low retrieved temperature, $424\pm33$~K.

\begin{figure}[t!]
\begin{center}
\includegraphics[width=0.7\textwidth]{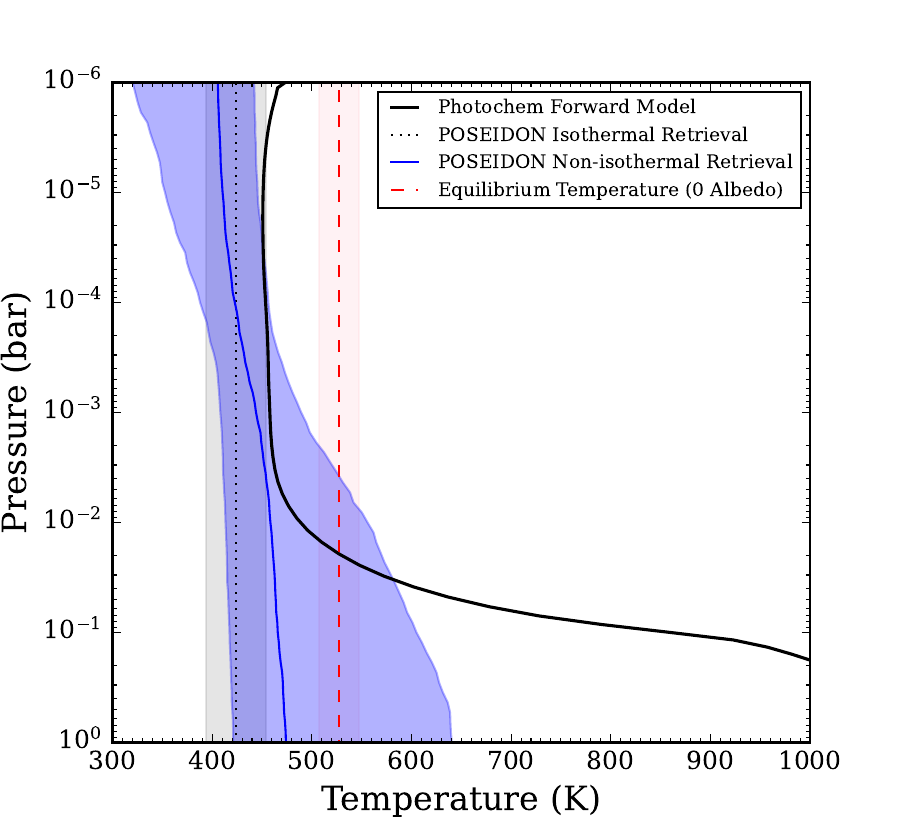}
\end{center}
\caption{{\bf Pressure-Temperature profile of V1298 Tau e.} The black curve shows the best-fit \texttt{PICASO} forward model, while the blue curve represents the non-isothermal profile retrieved with \texttt{POSEIDON} following the prescription of \cite{Madhusudhan2009}. The gray dotted line denotes the best-fit isothermal temperature in retrieval. Shaded regions indicate 1$\sigma$ confidence intervals. The forward model and retrieval show good agreement, whereas the zero-Bond-albedo equilibrium temperature (red curve) is significantly higher.}
\end{figure}

\end{document}